\documentclass[mathleft]{an}

\usepackage{graphicx}
\usepackage{times}
\begin{document}

\Pagespan{1}{}% Document's page range. 
% If second parameter is left empty, the last page is computed automatically.
\Yearpublication{2007}%
\Yearsubmission{2007}%
\Month{0}%   
\Volume{0}%  
\Issue{0}% 

\title{Correcting second-order contamination in low-resolution spectra}
\author{
V. Stanishev\inst{1}$^{\star\star}$
{\rm vall@physto.se}}

\institute{Department of Physics, Stockholm University, 
Albanova University Center, 106 91 Stockholm, Sweden}

\date{Received ; accepted }


\abstract{An empirical method for correcting low-resolution astronomical
spectra for second-order contamination is presented. The method was developed
for correcting spectra obtained with grism\,\#4 of the ALFOSC spectrograph 
at the {\it Nordic Optical Telescope} and the performance is
 demonstrated on spectra of two nearby bright Type Ia supernovae.}

\maketitle

\section{Introduction}

The availability of large format CCDs with high quantum efficiency over a wide
wavelength range, makes possible most modern low-resolution spectrographs to have the capability 
to obtain spectra that cover the whole optical range (3200-10000\AA) in a single exposure. 
However, it follows from the theory of diffraction gratings 
(see, e.g.,Schroeder  \cite{th}) that different diffraction orders overlap, i.e. a photon with
wavelength $\lambda^m$ in the $m$-th order will be diffracted at the same
direction as a photon with wavelength $\lambda^{m+1}$ from the $m+1$-st order and thus both will be
recorded at the same pixel on the detector. For diffraction 
gratings the relation between $\lambda^m$ and $\lambda^{m+1}$ is simple,
$\lambda^m=(m+1)\lambda^{m+1}/m$ (but see Gutierrez-Moreno et al. (\cite{guti})
for a case where this does not hold true due the a specific spectrograph design). 
Many spectrographs employ grisms instead of gratings, in which case the 
wavelength overlap relation is not that simple and is generally a non-linear function:
\begin{equation}
\lambda^{m}=f^{(m+1\rightarrow m)}(\lambda^{m+1}).
\label{eq:over}
\end{equation}
Most low-resolution astronomical spectra are obtained in the first diffraction 
order (with typical dispersion $\sim2-5$~\AA\,pixel$^{-1}$) and thus beyond
$\sim6000$\,\AA\ will be contaminated by the blue light of the 2nd-order, in many cases 
quite significantly.
The order contamination greatly compromises the 
spectrophotometric accuracy but also affects the measured equivalent 
widths and profiles of the lines, as well as may introduce
spurious lines when emission line objects are studied 
(Gutierrez-Moreno et al. \cite{guti}). The traditional way to
overcome this problem has been to use blue light blocking filters or two different 
spectrometer settings. In either case one either loses the blue part of the 
spectrum or doubles the exposure time. Double-beam spectrographs are another
solution, however, these are not as common instruments as the single-beam 
spectrographs. Besides, 
 the dichroics used to split the beam may have
non-uniform response at certain wavelengths, making the accurate flux
calibration challenging.  
It would be therefore beneficial in many cases if a methodology
for correcting {\it single} spectra for the higher-order contamination is 
available.  However, little has been done in that direction so far. 
In previous works Bowers et al. (\cite{bow}), P\'eroux et al. (\cite{per}) and 
 Norman at al. (\cite{nor})  only briefly outline the procedures they
use to correct the order contamination in their spectra, and 
Gutierrez-Moreno et al. (\cite{guti}), Leonard et al. (\cite{leo}) and 
 Steinmetz et al. (\cite{stain}) discuss other aspects of the order
contamination. 
I started working on the development of a 
method  for 2nd-order correction in early 2004, largely  motivated by the desire to 
optimize the Target-of-Opportunity observations 
of Type Ia Supernovae (SNe\,Ia) that began at the Nordic Optical Telescope  (NOT) 
in the fall of 2003 using the ALFOSC spectrograph. Because grism \#4 (3200-9100~\AA) suffers 
of severe 2nd-order  contamination beyond $\sim5800$\,\AA, 
observations with the red grism \#5 (5000-9800~\AA; 
free of order contamination) were also obtained,  
but this doubled the observing time. 
Eventually, in the end of 2004 Szokoly et al.
(\cite{szo}) published a paper where they also presented  in detail
a method for  correcting for the 2nd-order contamination; 
but my work was done  independently, albeit arriving at quite a similar method.

\section{The method}

 The {\it observed} spectrum in units 
of detected counts, $N(\lambda^{\rm I})$, is
$N(\lambda^{\rm I})=N^{\rm I}(\lambda^{\rm I})+N^{\rm II}(\lambda^{\rm I})$, 
where $N^{\rm I}(\lambda^{\rm I})$ is the {\it true} 1st-order spectrum that would be detected 
if there was no contamination
and $N^{\rm II}(\lambda^{\rm I})$ is the 2nd-order contamination, which is the 
true 2nd-order spectrum $N^{\rm II}(\lambda^{\rm II})$ mapped on the 
1st-order wavelengths $\lambda^{\rm I}$ with Eq.\,\ref{eq:over}.
The efficiency of a grism\footnote{Throughout the paper I will refer to grisms,
but the discussions are valid for diffraction grating as well.}  at a given order
should depend only on the parameters of the grim (Palmer \& Loewen \cite{gr})\footnote{The
efficiency of reflection gratings depend sensitively on the polarization 
of the incidence light. However, the transmission gratings, to
which class grisms belong, are practically free from the polarization effects 
(Palmer \& Loewen \cite{gr}).
Besides, in most cases the observed objects will not be strongly polarized.}. 
In this case it is easy to show that for any object observed, 
the 2nd-order photons with wavelength $\lambda^{\rm II}$ that are recorded 
at  wavelength $\lambda^{\rm I}$ in the first order, 
$N^{\rm II}(\lambda^{\rm I})$, can be 
expressed as a function of the detected 1st-order photons $N^{\rm
I}(\lambda^{\rm I})$:
\begin{equation}
N^{\rm II}(\lambda^{\rm I})=C(\lambda^{\rm I})N^{\rm I}(\lambda^{\rm I}).
\label{eq3}
\end{equation}
The function $C(\lambda^{\rm I})$ is 
the ratio of the efficiencies\footnote{efficiency here means the fraction of the 
incident photons in the corresponding order that are recorded on the detector.} 
of the 2nd- and 1st-order $C(\lambda^{\rm I})=E^{\rm II}(\lambda^{\rm I})/E^{\rm I}(\lambda^{\rm I})$,
where $E^{\rm II}(\lambda^{\rm I})$ is the efficiency of the 2nd-order again mapped on the 
1st-order wavelengths.
Thus the application of the method involves to steps: 1) wavelength 
transformation of the observed spectrum according to Eq.\,\ref{eq:over}, i.e.
$N(\lambda^{\rm I})\rightarrow N(f^{({\rm II}\rightarrow {\rm I})}(\lambda^{\rm I}))$ and 2) multiplying 
$N(f^{({\rm II}\rightarrow {\rm I})}(\lambda^{\rm I}))$ by  $C(\lambda^{\rm I})$ 
and subtracting it from $N(\lambda^{\rm I})$. What needs to be
determined is the wavelength relation  Eq.\,\ref{eq:over}  and $C(\lambda^{\rm I})$.

\begin{figure}[t]
\centering
\includegraphics*[width=8cm]{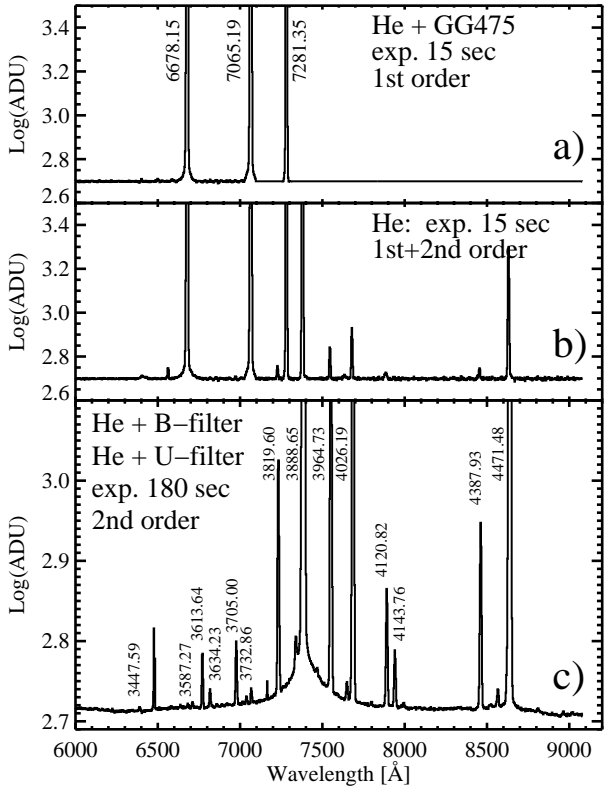}  %8cm
\caption{He arc-lamp spectra {\bf a)} with and 
 {\bf b)} without the blocking filter GG475. {\bf c)}
The combined spectrum obtained through the $U$ and $B$ filters. The lines
used to obtain the wavelength solution for the second order are labeled in 
the {\bf c)} panel. Some of these lines are clearly seen in the spectrum without
the blocking filter.}
\label{f:he}
\end{figure}

The work on the method was first initiated by the observations of SN\,2004S
(Krisciunas et al. \cite{kri_04s}) that were obtained  at NOT
in February 2004 using ALFOSC plus grism~\#4. 
Because of the  $\sim-31^\circ$ declination, the observations at La Palma 
were performed at very high airmass, often $X>2$,
with slit along the parallactic angle to minimize the slit losses due
to the differential atmospheric refraction (Filippenko \cite{diff}). 
At such a high airmass,  the   
atmospheric refraction is large and the atmosphere effectively  acts as a
cross-disperser, similar to cross-dispersers in 
the Echelle spectrographs. This allowed on one occasion
the 2nd-order to be spatially separated from the first order 
(see Fig.\,\ref{f:t2}, top), and the  spectra at the two
orders could be individually extracted. 
My first attempt was to use this observation to
derive a method for correcting  for the contamination. With too few blue photons
detected (because of the high airmass) this was unsuccessful, but stimulated 
further work using observations of bright, blue stars with and without
order-blocking filters in order to isolate the 2nd-order light.
 $N^{\rm II}(\lambda^{\rm I})$ is the 
difference between the photon flux detected without and with the
filter.
Such observations were only obtained in November 2005. Using a slightly 
adjusted theoretical wavelength overlap function for grism \#4 provided by Per Rasmussen 
(Copenhagen Observatory, private communication) the method was 
successfully tested. However, more observations were obtained in May 2006,
including dedicated observations to determine experimentally the wavelength
overlap relation Eq.\,\ref{eq:over}. The results presented here are based on
this later observing run.

\begin{figure}[t]
\centering
\includegraphics*[width=8cm]{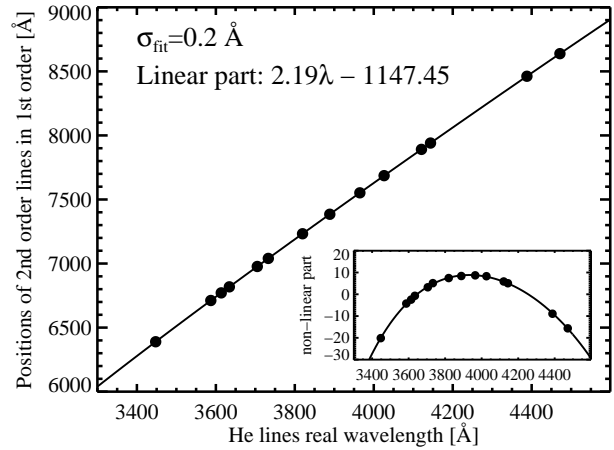}  %8cm
\caption{The overlap relation between the 1st and 2nd order.
The 1$\sigma$ scatter around the fit is 0.2\AA.
{\bf Inset:} the non-linear part of the fit.}
\label{f:over}
\end{figure}

To determine the overlap relation Eq.\,\ref{eq:over} observations of 
Ne and He arc-lamps were obtained. The 1st-order wavelength solution was
obtained from exposure of the He and Ne lamps simultaneously. Four 3 min 
long  exposures of the He lamp without any filter and through the
$U$, $B$, and the filter GG475 that block the light shortward of $\sim4500$\,\AA\
were also obtained. 
The $U$ and  $B$ filters served to
block the 1st-order He lines, so that the faint 2nd-order lines cold be
measured to derive the wavelength solution for the 2nd-order
(Fig.\,\ref{f:he}). The other two exposures were only used to check if the
filters introduced shifts to the line positions. Shifts of 1-2~pixels were found (and
corrected) for the $U$ and  $B$ filters, and none for GG475. 
Figure\,\ref{f:over} shows the positions of the 2nd-order He 
lines as detected in the 1st-order vs. their real wavelengths. 
These data determine the order overlap relation
and as can be seen
from the inset in Fig.\,\ref{f:over} it is non-linear.
Note also that because of the higher dispersion,  more lines could be 
detected in the  
second order than are usually seen in the
first order.

\begin{figure}[t]
\centering
\includegraphics*[width=8cm]{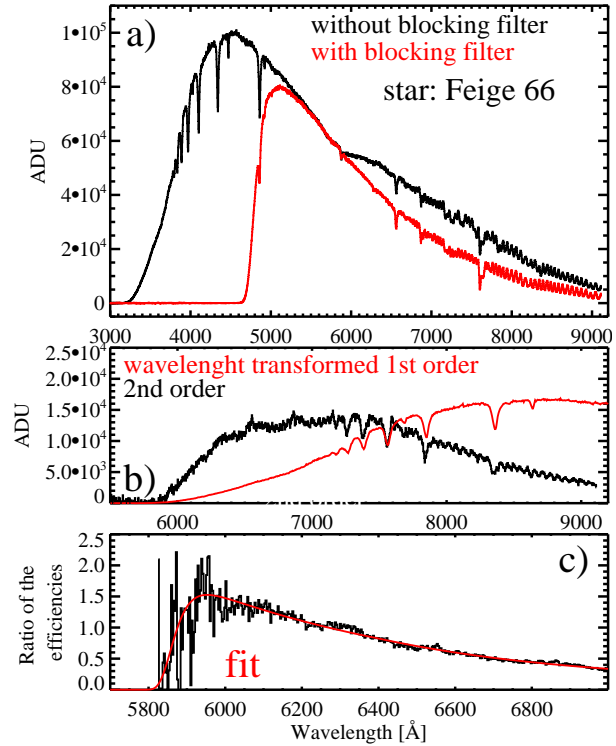}  %8cm
\caption{{\bf a)} The spectra of the hot sdO star
Feige 66 with and without the 
GG475 blocking filter. {\bf b)} The 2nd-order light and the wavelength
transformed 1st-order light. {\bf c)} The ratio between the 2nd and 
1st-order light. The parametrized fit which is used for 
the 2nd-order contamination correction is also shown.}
\label{f:std}
\end{figure}

Figure\,\ref{f:std}a shows the observations of the spectrophotometric standard
Feige 66 with and without the  GG475 filter.  Figure\,\ref{f:std}b
shows the difference between the two spectra 
($=N^{\rm II}(\lambda^{\rm I})$)\footnote{The GG475 
filter transmission decreases nearly linearly 
from 97\% to 92\% between 5800\,\AA\ and 9000\,\AA. The flux observed through
the blocking filter was corrected accordingly. This correction 
was not in the original method and was included later after a note  
in Szokoly et al. (\cite{szo}) that the blue light blocking filter may 
also affect the red part of the spectrum. This is the only influence that the 
paper of Szokoly et al. had  on the work presented here.}
 and the wavelength
transformed observed spectrum 
$N(\lambda^{\rm I})\rightarrow N(f^{({\rm II}\rightarrow {\rm I})}(\lambda^{\rm I}))$. 
Note that the
Hydrogen Balmer lines from the 2nd-order are clearly seen in the difference spectrum and their
position  are well matched by the lines in the transformed spectrum.
Figure\,\ref{f:std}c shows the ratio 
$N^{\rm II}(\lambda^{\rm I})/N(f^{({\rm II}\rightarrow {\rm I})}(\lambda^{\rm I})$ which is
the correction function $C(\lambda^{\rm I})$. 
Because of the low sensitivity in the blue,
the part between 5800\AA\ and 6000\AA\ is rather noisy and I fit a parametrized
function (the thick line), which is the final correction function $C(\lambda^{\rm I})$. 

\section{Performance}

To demonstrate the performance of the algorithm I use observations of two
Type Ia supernovae (SNe)
 2005cf (Garavini et al. \cite{gar}) and 2005hk 
(Stanishev et al. \cite{stan}). Both SNe were observed at NOT with grisms 
\#4 and \#5 as a part of large observing campaigns involving many other 
telescopes. If there were no order contamination in grism \#4,
the shape of the flux calibrated spectra with the two grisms would be the same.
I therefore correct the grism \#4 spectra for the 2nd-order
contamination and compare them with the ones obtained with
grism \#5  in order to evaluate the performance of the algorithm. Note that 
the observing conditions were probably not perfectly photometric and so there is
difference in the {\it absolute} flux level of the grism \#4 and \#5 spectra of the order 
of $\sim5-15$\%. Such small variations should not afect the {\it relative} flux 
calibration and are not problem for the analysis presented here. 

\begin{figure}[t]
\centering
\includegraphics*[width=8cm]{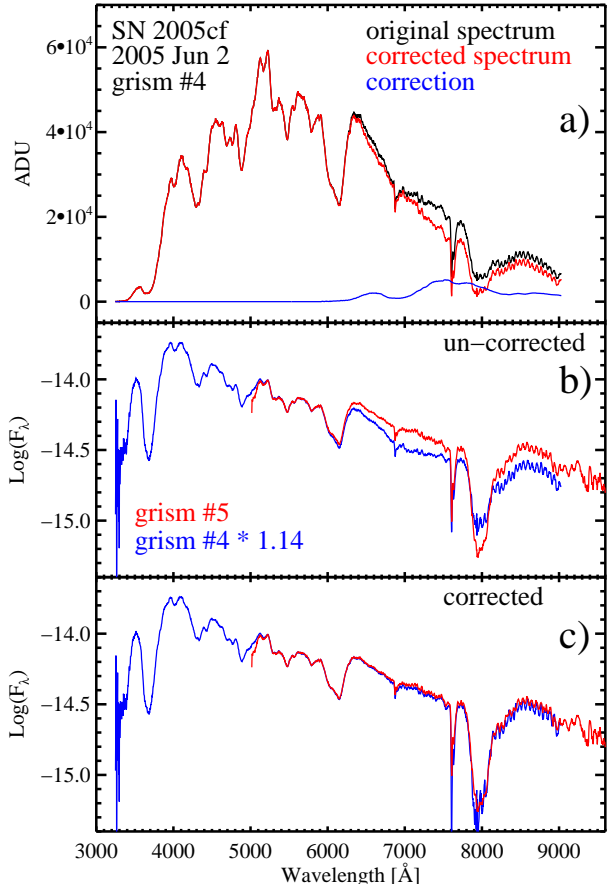}  %8cm
\caption{{\bf a)} The observed, the calculated correction and the corrected 
 spectra of SN 2005cf observed at NOT on 2005 June 2.{\bf b)} The flux calibrated 
{\it uncorrected} spectrum with grism \#4 compared with the spectrum 
 with grism \#5. The effect of the 2nd-order contamination in grism \#4 is 
very strong. {\bf c)} Comparison of the two spectra after the 
one with grism \#4 has been corrected for the 2nd-order contamination. The
two spectra match almost perfectly.}
\label{f:t1}
\end{figure}

Figure\,\ref{f:t1} presents the results for SN 2005cf. In Fig.\,\ref{f:t1}a
are shown the observed spectrum with grism \#4, the estimated correction
and the corrected spectrum. Figure\,\ref{f:t1}b shows the uncorrected, 
flux calibrated spectra with the two grisms. Note that for this example neither
the SN nor the spectrum of the spectrophotometric standard BD~+33~2642 used for the
flux calibration were corrected. Because BD~+33~2642 is actually bluer than the
SN, effectively flux is {\it subtracted} from and not {\it added} to the SN
spectrum. After correcting for the 2nd-order contamination, the spectra with
the two grisms line up nicely (Fig.\,\ref{f:t1}c). Figure\,\ref{f:t1}b clearly
demonstrates not only the problem with the accurate relative flux calibration, but
also the negative effect on the measured line equivalent widths: 
in the uncorrected grism \#4 spectrum the strong Ca~II infrared triplet at
$\sim8000$\AA\ is much weaker than in the grism \#5 spectrum.

\begin{figure}[t]
\centering
\includegraphics*[width=8cm]{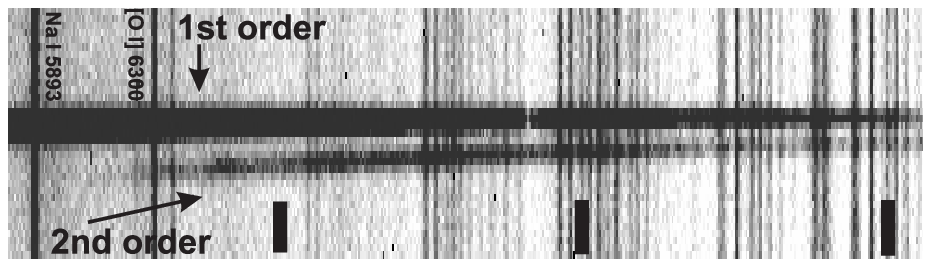} \\
\includegraphics*[width=7.5cm]{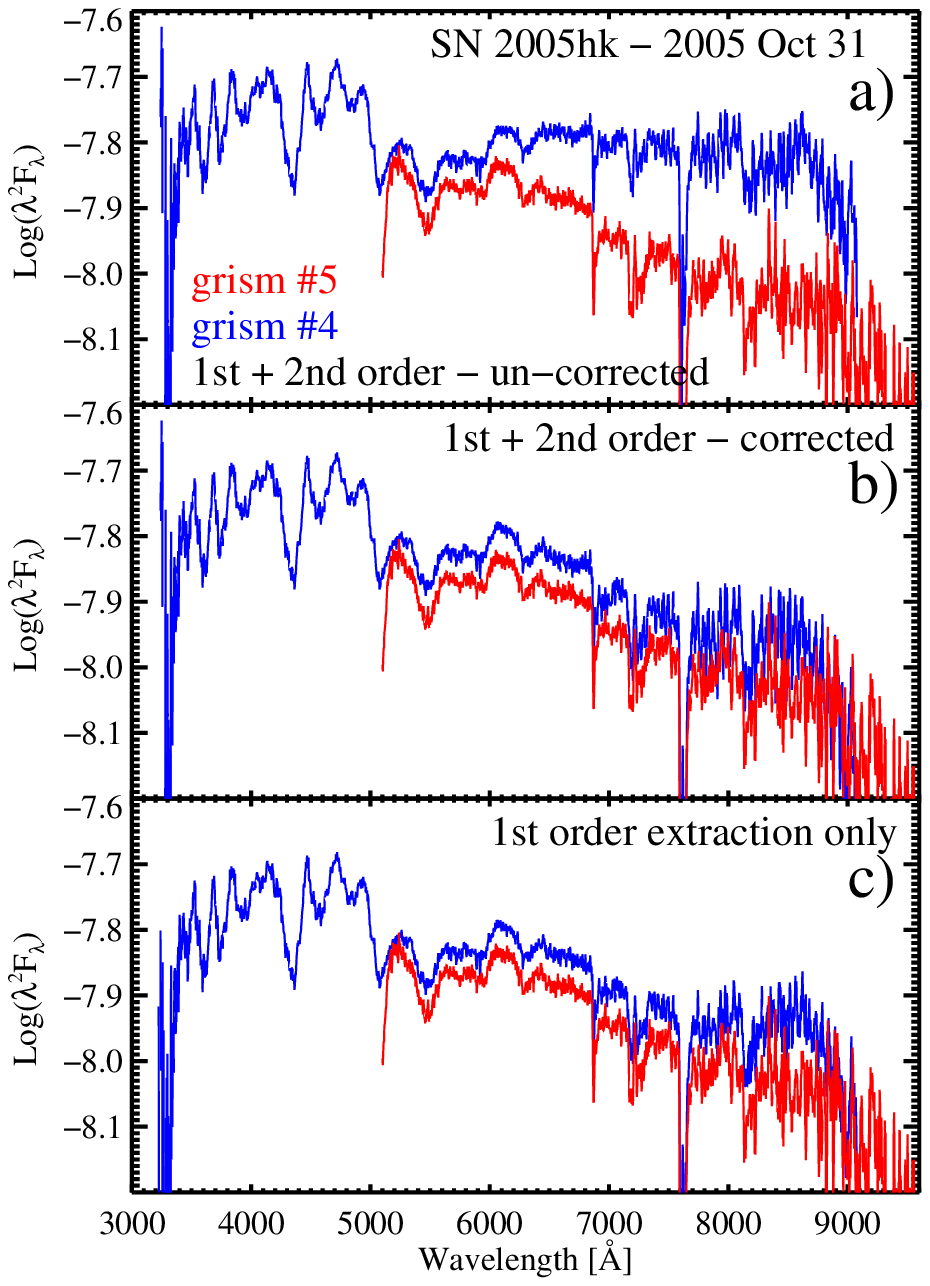} 
\caption{{\bf Upper panel:} The two-dimensional spectral image
of SN 2005hk observed at NOT on 2005 October 31 with grism \#4. The
1st and 2nd orders are clearly seen spatially separated (see the text for 
details). {\bf a)}  The flux calibrated {\it uncorrected} spectrum with 
grism \#4 compared with the spectrum   with grism \#5.
 {\bf b)} Comparison of the two spectra after the 
one with grism \#4 has been corrected for the 2nd-order contamination.
{\bf c)} Comparison between the grism \#5 spectrum with the uncorrected 
grism \#4 spectrum. In this case a narrow extraction aperture was used 
for grism \#4 spectrum, so that only the 1st order light was extracted.
}
\label{f:t2}
\end{figure}

 With respect to SN~2005hk, I first correct the spectrophotometric standards
to derive a contamination-free response function, 
which is then used to flux calibrate the SN spectra. On  2005 October 31
SN~2005hk was observed at high airmass and similarly to the case of the SN~2004S
observations described in the previous section, the second order can be seen
spatially well separated from the first order (Fig.\,\ref{f:t2} top). 
The three ticks in Fig.\,\ref{f:t2} (top) roughly correspond to wavelengths of 
$\sim$6700, 7900, and 9000\,\AA\ from left to right. Cuts along the spatial axis 
were taken at these three positions in order to measure the FWHM of the 
spectral traces and the separation between them. The separations  
are 12, 10 and 7 pixels from blue to red wavelengths (1 pixel=0.19 arcsec). The FWHM of the 
1st-order trace is 4.6\,pixles at all cuts, but the second order trace FWHM 
decreases from 6 to 5 pixels from blue to red.  The large separation between the orders
 allows to extract the 1st-order 
light with almost no 2nd-order contamination and to compare how the 
correction works not only with the grism \#5 spectrum but also with this
'clean' 1st-order extraction. I used an extraction aperture of 9 pixels width
and thus below $\sim8000$\,\AA\ the extracted spectrum should be to a
large extent  
contamination-free, while above $\sim8000$\,\AA\ the contaminations will be somehow larger
because the separation between the traces decreases to the red. Nonetheless, I find that
the so extracted grism \#4 spectrum is in a good agreement with the grism 
\#5 one, indicating that nearly 'clean' extraction have been  achieved.
The results are presented in Fig.\,\ref{f:t2}
and turning our attention to the 2nd-order contamination correction, I again find that 
the contamination has been very well corrected.

It should be noted that the experience from the observations shows 
that separation between the orders as large as 10 pixels
is very rare. It is normally of the order of few pixels only and the two 
orders are blended together\footnote{large seeing will cause blending even in
case of large separations}. For the algorithm to work correctly, the spectrum that 
is to be corrected has to contain the entire light from both, 
the 1st and the 2nd order. Given the possibility 
for a significant separation between the orders, to make sure that this is 
indeed the case one should carefully select the
aperture position and width so that the light in both orders is fully extracted.

\section{Conclusions}

I present an empirical method for correcting low-resolution astronomical 
spectra for 2nd-order contamination. The method was applied to grism \#4 of the 
ALFOSC instrument at the Nordic Optical Telescope, but can be used at  
any low-resolution grating/grism spectrograph. The performance of the 
method was tested on observations of two bright nearby Type Ia supernovae,
SNe 2005cf and 2005hk, and in both cases the result was excellent, greatly 
improving the accuracy of the final flux calibrated spectra. Thus,
in many situations the method would allow to reduce almost twice the 
observing time of programs that need to observe the whole optical range. 
Various Target-of-Opportunity programs on Supernovae,  Gamma Ray Bursts and
other transient object can greatly benefit from it. I expect the method to work
very well for low-polarized objects with broad spectral features like SNe and GRBs. 
Its performance on objects with narrow spectral features and/or high polarization
should be further investigated. 

\begin{acknowledgements}
This work is supported in part by the European Community's Human
Potential Program ``The Physics of Type Ia Supernovae'', under
contract HPRN-CT-2002-00303.  The author would like to thank
the G\"oran Gustafsson Foundation for financial support. I am grateful 
to Jakob J\"onson and Amanda Djupvik for performing the observation. 
ALFOSC  is 
owned by the Instituto de Astrofisica de Andalucia (IAA) and 
operated at the Nordic Optical Telescope under agreement between 
IAA and the NBIfAFG of the Astronomical Observatory of Copenhagen.
\end{acknowledgements}


\begin{thebibliography}{}


\bibitem[1997]{bow} Bowers, E.~J.~C., Meikle, W.~P.~S., Geballe, T.~R., 
Walton, N.~A., Pinto, P.~A., Dhillon, V.~S.: 1997, MNRAS 290, 663

\bibitem[1982]{diff} Filippenko, A.~V.: 1982, PASP 94, 715

\bibitem[2007]{gar} Garavini, G., Nobili, S., S. Taubenberger, S., et al.:
2007, A\&A, accepted (astro-ph/0702569 )

\bibitem[1994]{guti} Gutierrez-Moreno, A., Heathcote, S., Moreno, H., Hamuy, M.:
1994, PASP 106, 1184

\bibitem[2002]{leo} Leonard, D.~C., Filippenko, A.~V., Chornock, R., 
Foley, R.~J.: 2002, PASP 114, 1333

\bibitem[2007]{kri_04s} Krisciunas, K.,  Garnavich, P.~M., Stanishev, V., et al.:
2007, AJ 133, 58

\bibitem[2002]{nor} Norman, C., Hasinger, G., Giacconi, R., et al.: 
2002, ApJ 571, 218

\bibitem[2006]{gr} Palmer, Ch. \& Loewen, E.: 2005,  
{\it Diffraction grating Handbook -- 6th edition}, Newport Corporation, 
available at {\tt http://gratings.newport.com/library/ handbook/handbook.pdf}

\bibitem[2001]{per} P\'eroux, C., Storrie-Lombardi, L.~J., McMahon, R.~G., Irwin, M.,
Hook, I.~M.: 2001, AJ 121, 1799

\bibitem[2000]{th} Schroeder, D.J.: 2000, {\it Astronomical optics}, 2nd ed.

\bibitem[2006]{stain} Steinmetz, M., Zwitter, T., Siebert, A., et al.: 2006, 
AJ 132, 1645

\bibitem[2004]{szo} Szokoly, G. P., Bergeron, J., Hasinger, G., et al.:   2004, ApJS 155, 271

\bibitem[2007]{stan} Stanishev, V., et al.: 2007, in preparation


\end{thebibliography}
\end{document}